\documentclass[twoside,11pt]{10ssmmp} 
\usepackage{fancyhdr}
\usepackage{amsfonts}

\usepackage{xcoffins}
\NewCoffin\tablecoffin

\usepackage{amsmath,amssymb}

\newcommand{\lc}{\varepsilon}

\newcommand{\ds}{\displaystyle}
\newcommand{\del}{\partial}
\newcommand{\itGamma}{{\mit{\Gamma}}}
\newcommand{\D}{\mathrm{d}}

\newcommand{\diag}{\mathop{\rm diag}\nolimits}
\newcommand{\rmd}{\mathrm{d}}
\newcommand{\killing}[2]{ {\langle #1 \, , #2 \rangle } }
\newcommand{\komut}[2]{ { [ \,#1\, ,\, #2\, ] } }

\newcommand{\realni}{\ensuremath{\mathbb{R}}}

\newcommand{\grasmanovi}{\ensuremath{\mathbb{G}}}

\newcommand{\cA}{{\cal A}}
\newcommand{\cD}{{\cal D}}
\newcommand{\cF}{{\cal F}}
\newcommand{\cG}{{\cal G}}
\newcommand{\cH}{{\cal H}}

\newcommand{\cM}{{\cal M}}

\begin{document}

\setcounter{page}{251}


\title{Construction and examples of higher gauge theories\hspace{.25mm}\thanks{\,This work was supported by the project ON171031 of the Ministry of Education, Science and Technological Development (MPNTR) of the Republic of Serbia, and partially by the bilateral scientific cooperation between Austria and Serbia through the project ``Causality in Quantum Mechanics and Quantum Gravity - 2018-2019'', no. 451-03-02141/2017-09/02, supported by the Federal Ministry of Science, Research and Economy (BMWFW) of the Republic of Austria, and the Ministry of Education, Science and Technological Development (MPNTR) of the Republic of Serbia.}}


\author{
\bf{Tijana Radenkovi\'c}\hspace{.25mm}\thanks{\,e-mail address: rtijana@ipb.ac.rs} \\
\normalsize{Institute of Physics, University of Belgrade,} \\
\normalsize{Pregrevica 118, 11080 Belgrade, Serbia} \vspace{2mm} \\
\bf{Marko Vojinovi\'c}\hspace{.25mm}\thanks{\,e-mail address: vmarko@ipb.ac.rs} \\
\normalsize{Institute of Physics, University of Belgrade,} \\
\normalsize{Pregrevica 118, 11080 Belgrade, Serbia}
}

\date{} 

\maketitle 

\begin{abstract}
We provide several examples of higher gauge theories, constructed as generalizations of a $BF$ model to $2BF$ and $3BF$ models with constraints. Using the framework of higher category theory, we introduce appropriate $2$-groups and $3$-groups, and construct the actions for the corresponding constrained $2BF$ and $3BF$ theories. In this way, we can construct actions which describe the correct dynamics of Yang-Mills, Klein-Gordon, Dirac, Weyl, and Majorana fields coupled to Einstein-Cartan gravity. Each action is naturally split into a topological sector and a sector with simplicity constraints. The properties of the higher gauge group structure opens up a possibility of a nontrivial unification of all fields.
\end{abstract}

\section{\label{secI}Introduction}

The quantization of the gravitational field is one of the fundamental open problems in modern physics. There are various approaches to this problem, some of which have developed into vast research frameworks. One of such frameworks is the Loop Quantum Gravity approach, which aims to establish a nonperturbative quantization of gravity, both canonically and covariantly \cite{RovelliBook, RovelliVidottoBook, Thiemann2007}. The covariant approach is slightly more general, and focuses on providing a possible rigorous definition of the path integral for the gravitational field,
\begin{equation} \label{StatementOfIntent}
Z = \int \cD g\; e^{iS[g]}\,.
\end{equation}
This is done by considering a triangulation of a spacetime manifold, and defining the path integral as a discrete state sum of the gravitational field configurations living on the simplices in the triangulation. This quantization technique is known as the {\em spinfoam} quantization method, and is performed via the following three steps:
\begin{enumerate}
\item[(1)] one writes the classical action $S[g]$ as a constrained $BF$ action;
\item[(2)] one uses the Lie group structure, underlying the topological sector of the action, to define a triangulation-independent state sum $Z$;
\item[(3)] one imposes the simplicity constraints on the state sum, promoting it into a triangulation-dependent state sum, which serves as a definition for the path integral (\ref{StatementOfIntent}).
\end{enumerate}
So far, this quantization prescription has been implemented for various choices of the gravitational action, of the Lie group, and of the spacetime dimension. For example, in $3$ dimensions, historically the first spinfoam model is known as the Ponzano-Regge model \cite{PonzanoRegge1968}. In $4$ dimensions there are multiple models, depending on the choice of the Lie group and the way one imposes the simplicity constraints \cite{BarrettCrane,BarrettCrane1,Ooguri,EPRL,FK}. While these models do give a definition for the gravitational path integral, none of them are able to consistently include matter fields. Including the matter fields has so far had limited success \cite{RovelliSpinfoamFermions}, mainly due to the absence of the tetrad fields from the topological sector of the theory.

In order to resolve this issue, a new approach has been developed, using the framework of {\em higher gauge theory} (see \cite{BaezHuerta2011} for a review). In particular, one uses the idea of a {\em categorical ladder} to generalize the $BF$ action (based on a Lie group) into a $2BF$ action (based on the so-called $2$-group structure). A suitable choice of the {\em Poincar\'e $2$-group} introduces the needed tetrad fields into the topological sector of the action \cite{MikovicVojinovic2012}. While this result opened up a possibility to couple matter fields to gravity, the matter fields could not be naturally expressed using the underlying algebraic structure of a $2$-group, rendering the spinfoam quantization method inapplicable. Namely, the matter sector could indeed be added to the classical action, but could not be expressed itself as a constrained $2BF$ theory, which means that the steps 1--3 above could not be performed for the matter sector of the action, but only for gravity.

This final issue has recently been resolved in \cite{RadenkovicVojinovic2019}, by passing from the $2$-group structure to the $3$-group structure, generalizing the action one step further in the categorical ladder. This generalization naturally gives rise to the so-called $3BF$ action, which turns out to be suitable for a unified description of both gravity and matter fields. The steps of the categorical ladder and their corresponding structures are summarized as follows:

{\footnotesize
\begin{center}
\setlength{\tabcolsep}{0pt}
     \label{tab:table5}\begin{tabular}{|c|c|c|c|c|} \hline
\shortstack{categorical$\vphantom{A^A}$ \\ structure} & \shortstack{algebraic$\vphantom{A^A}$ \\ structure} & \shortstack{linear$\vphantom{gA^A}$ \\ structure} & \shortstack{topological$\vphantom{A^A}$ \\ action} & \shortstack{degrees of$\vphantom{A^A}$ \\ freedom} \\ \hline\hline
Lie group$\vphantom{\ds\int}$ & Lie group & Lie algebra & $BF$ theory & gauge fields \\ \hline
Lie $2$-group$\vphantom{\ds\int}$ & \raisebox{-0.5em}[1em][0em]{\shortstack{Lie crossed\\ module}} & \raisebox{-0.5em}[1em][0em]{\shortstack{differential Lie \\ crossed module}} & $2BF$ theory & tetrad fields \\ \hline
\ \ Lie $3$-group$\vphantom{\ds\int}$\ \ \  &\ \  \raisebox{-0.5em}[1em][0em]{\shortstack{Lie $2$-crossed\\ module}}\ \ \  &\ \  \raisebox{-0.5em}[1em][0em]{\shortstack{differential Lie \\ $2$-crossed module}}\ \ \  &\ \  $3BF$ theory\ \ \  & \ \ \raisebox{-0.5em}[1em][0em]{\shortstack{scalar and \\ fermion fields}}\ \ \  \\ \hline
\end{tabular}
\end{center}
}

The purpose of this paper is to give a systematic overview of the constructions of classical $BF$, $2BF$ and $3BF$ actions, both pure and constrained, in order to demonstrate the categorical ladder procedure and the construction of higher gauge theories. In other words, we focus on the step $1$ of the spinfoam quantization programme.

The layout of the paper is as follows. Section 2 deals with models based on a $BF$ theory. First we discuss the pure, topological $BF$ theory, and then pass on to the the physically more interesting Yang-Mills theory in Minkowski spacetime and the Plebanski formulation of general relativity. In Section 3 we study the first step in the categorical ladder, namely models based on the $2BF$ theory. After introducing the pure $2BF$ theory, we study the relevant formulation of general relativity \cite{MikovicVojinovic2012}, and then the coupled Einstein-Yang-Mills theory. Then, in Section 4 we perform the second step in the categorical ladder, passing on to models based on the $3BF$ theory. After the introduction of the pure $3BF$ model, we construct constrained $3BF$ actions for the cases of Klein-Gordon, Dirac, Weyl and Majorana fields, all coupled to the Einstein-Cartan gravity in the standard way. As we shall see, the scalar and fermion fields will be {\em naturally associated to a new gauge group}, generalizing the purpose of a gauge group in the Yang-Mills theory, which opens up a possibility of an algebraic classification of matter fields. Finally, Section 5 contains a discussion and conclusions.

The notation and conventions are as follows. The local Lorentz indices are denoted by the Latin letters $a,b,c,\dots$, take values $0,1,2,3$, and are raised and lowered using the Minkowski metric $\eta_{ab}$ with signature $(-,+,+,+)$. Spacetime indices are denoted by the Greek letters $\mu,\nu,\dots$, and are raised and lowered by the spacetime metric $g_{\mu\nu} = \eta_{ab} e^a{}_{\mu} e^b{}_{\nu}$, where $e^a{}_{\mu}$ are the tetrad fields. The inverse tetrad is denoted as $e^{\mu}{}_a$. All other indices that appear in the paper are dependent on the context, and their usage is explicitly defined in the text where they appear. We work in the natural system of units where $c=\hbar=1$, and $G = l_p^2$, where $l_p$ is the Planck length.

\section{\label{secII}$BF$ theory}

We begin with a short review of $BF$ theories. See \cite{BFgravity2016, zakopane, plebanski1977} for additional information.

\subsection{\label{secIIsi}Pure $BF$ theory}

Given a Lie group $G$, and denoting its corresponding Lie algebra as $\mathfrak{g}$, one introduces the pure $BF$ action as follows (we limit ourselves to the physically relevant case of $4$-dimensional spacetime manifolds $\cM_4$):
\begin{equation}\label{eq:bf}
    S_{BF} =\int_{\cM_4} \langle B \wedge \cF \rangle_\mathfrak{g}\,.
\end{equation}
Here, $\cF\equiv \D \alpha+ \alpha\wedge \alpha$ is the curvature $2$-form for the algebra-valued connection $1$-form $\alpha \in \cA^1(\cM_4\,, \mathfrak{g})$, and $B \in \cA^2(\cM_4\,, \mathfrak{g})$ is a Lagrange multiplier $2$-form, while $\langle \_\,,\_\rangle{}_{\mathfrak{g}}$ denotes a $G$-invariant bilinear symmetric nondegenerate form.

One can see from (\ref{eq:bf}) that the action is diffeomorphism invariant, and it is also gauge invariant with respect to $G$, provided that $B$ transforms as a scalar with respect to $G$.

Varying the action (\ref{eq:bf}) with respect to $B^{\beta}$ and $\alpha^{\beta}$, where the index $\beta$ is the group $G$ index (which counts the generators of $\mathfrak{g}$), one obtains the following equations of motion,
\begin{equation}
    \cF^{\beta}=0\,,\quad \quad \nabla B^{\beta} \equiv \D B^{\beta} + f_{\gamma\delta}{}^{\beta} \alpha^{\gamma} \wedge B^{\delta}  =0\,,
\end{equation}
where $f_{\gamma\delta}{}^{\beta}$ are the structure constants of the Lie group $G$. From the first equation of motion, one immediately sees that $\alpha$ is a flat connection, meaning that $\alpha=0$ up to gauge transformations. Given this, the second equation of motion implies that $B$ is constant. Therefore, there are no local propagating degrees of freedom, and the theory is called {\em topological}.

\subsection{\label{SecIIsii}Yang-Mills theory}

In physics one is usually interested in theories which are not topological, i.e., which have local propagating degrees of freedom. As a rule of thumb, one recognizes that the theory does have local propagating degrees of freedom if one of the equations of motion is a second-order partial differential equation, usually featuring a D'Alambertian operator $\Box$ in some form. In order to transform the pure $BF$ action into such a theory, one adds an additional term to the action, commonly called the {\em simplicity constraint}. The resulting action is called a {\em constrained $BF$ theory}. A nice example is the Yang-Mills theory for the $SU(N)$ group in Minkowski spacetime, which can be rewritten as a constrained $BF$ theory in the following way:
\begin{equation}\label{eq:bfgauge}
\begin{array}{lcl}
    S & = & \ds \int B_I\wedge F^I+ \lambda^I\wedge \Big(B_I-\frac{12}{g}{M_{ab}}_I\delta^a\wedge \delta^b \Big) \\
    & & \phantom{\ds\int} \ds  +\zeta^{ab}{}^I \Big( {M_{ab}}{}_I\varepsilon_{cdef}\delta^c\wedge \delta^d \wedge \delta^e \wedge \delta^f- g_{IJ}F^J \wedge \delta_a \wedge \delta_b \Big) \,. \\
 \end{array}
\end{equation}
Here $F \equiv \D A + A\wedge A$ is again the curvature $2$-form for the connection $A \in \cA^1(\cM_4\,, \mathfrak{su}(N))$, and $B\in \cA^2(\cM_4\,, \mathfrak{su}(N))$ is the Lagrange multiplier $2$-form. The Killing form $g_{IJ} \equiv \killing{\tau_I}{\tau_J}_{\mathfrak{su}(N)} \propto f_{IK}{}^L f_{JL}{}^K$ is used to raise and lower the indices $I,J,\dots$ which count the generators of $SU(N)$, while $f{}_{IJ}{}^K$ are the structure constants for the $\mathfrak{su}(N)$ algebra. In addition to the topological $B\wedge F$ term, there are also two simplicity constraint terms present, featuring two Lagrange multipliers, a $2$-form $\lambda^I$ and a $0$-form $\zeta^{abI}$. The $0$-form $M_{abI}$ is also a Lagrange multiplier, while $g$ is the coupling constant for the Yang-Mills theory.

Finally, $\delta^a$ is a nondynamical $1$-form, such that there exists a global coordinate frame in which its components are equal to the Kronecker symbol $\delta^a{}_{\mu}$ (hence the notation $\delta^a$). The 1-form $\delta^a$ plays the role of a background field, and defines the global spacetime metric, via the equation
\begin{equation} \label{eq:flatspacetimemetric}
\eta_{\mu\nu} = \eta_{ab} \delta^a{}_{\mu} \delta^b{}_{\nu}\,,
\end{equation}
where $\eta_{ab} \equiv \diag (-1,+1,+1,+1)$ is the Minkowski metric. Since the coordinate system is global, the spacetime manifold $\cM_4$ is understood to be flat. The indices $a,b,\dots$ are local Lorentz indices, taking values $0,\dots,3$. Note that the field $\delta^a$ has all the properties of the tetrad $1$-form $e^a$ in the flat Minkowski spacetime. Also note that the action (\ref{eq:bfgauge}) is manifestly diffeomorphism invariant and gauge invariant with respect to $SU(N)$, but not background independent, due to the presence of $\delta^a$.

Varying the action (\ref{eq:bfgauge}) with respect to the variables ${\zeta^{ab}}{}^I$, ${M_{ab}}{}_I$, $A^I$,  $B_I$, and $\lambda^I$, respectively (but not with respect to the background field $\delta^a$), we obtain the equations of motion:
\begin{gather}
\label{eq:g1}{M_{ab}}_I\varepsilon_{cdef}\delta^c \wedge \delta^d \wedge \delta^e \wedge \delta^f- F_I \wedge \delta_a \wedge \delta_b=0\,, \vphantom{\ds\int} \\
\label{eq:g2}-\frac{12}{g}\lambda^I\wedge \delta^a \wedge \delta^b + \zeta^{ab}{}^I\varepsilon_{cdef}\delta^c \wedge \delta^d \wedge \delta^e \wedge \delta^f=0\,, \vphantom{\ds\int} \\
\label{eq:g3}
-\D B_I+{f}_{JI}{}^K B_K\wedge A^J+\D(\zeta^{ab}{}_I \delta_a \wedge \delta_b)-{f}_{JI}{}^K \zeta^{ab}{}_K \delta_a \wedge \delta_b \wedge A^J=0\,, \vphantom{\ds\int} \\
\label{eq:g4}F_I+\lambda_I=0\,, \vphantom{\ds\int} \\
\label{eq:g5}B_I-\frac{12}{g}{M_{ab}}_I\delta^a\wedge \delta^b=0\,, \vphantom{\ds\int}
\end{gather}
From the equations (\ref{eq:g1}), (\ref{eq:g2}), (\ref{eq:g4}) and (\ref{eq:g5}) one obtains the multipliers as algebraic functions of the field strength $F^I{}_{\mu\nu}$ for the dynamical field $A^I$:
\begin{equation}
\begin{array}{ccc}
  \ds  M_{ab}{}_I=\frac{1}{48}\varepsilon_{abcd}F{}_I{}^{cd}\,, & \quad\vphantom{\ds\int} & \ds {\zeta^{ab}}{}^I=\frac{1}{4g}\varepsilon^{abcd}F{}^I{}_{cd}\,, \\
  \ds  \lambda{}_I{}_{ab}=F{}_I{}_{ab}\,, & \quad\vphantom{\ds\int^A} & \ds B{}_I{}_{ab}=\frac{1}{2g}\varepsilon_{abcd}F{}_I{}^{cd}\,. \\
\end{array}
\end{equation}
Here we used the notation $F_I{}_{ab}=F_I{}_{\mu\nu}\delta_a{}^\mu \delta_b{}^\nu$, and similarly for other variables, where we exploited the fact that $\delta^a{}_\mu$ is invertible. Using these equations and the differential equation (\ref{eq:g3}) one obtains the equation of motion for gauge field $A^I{}_{\mu}$,
\begin{equation}\label{eq:g8}
    \nabla_\rho F^{I\rho \mu}\equiv \partial_\rho F^{I\rho \mu} + f_{JK}{}^I A^J{}_\rho F^K{}^{\rho\mu}=0\,.
\end{equation}
This is precisely the classical equation of motion for the free Yang-Mills theory. Note that this is a second-order partial differential equation for the field $A^I{}_{\mu}$, and moreover contains the $\Box$ operator in the first term.

In addition to the Yang-Mills theory, one can easily extend the action (\ref{eq:bfgauge}) in order to describe the massive vector field and obtain the Proca equation of motion. This is done by adding a mass term
\begin{equation}
-\frac{1}{4!} m^2 A_{I\mu} A^I{}_{\nu} \eta^{\mu\nu} \lc_{abcd} \delta^a \wedge \delta^b \wedge \delta^c \wedge \delta^d
\end{equation}
to the action (\ref{eq:bfgauge}). Of course, this term explicitly breaks the $SU(N)$ gauge symmetry of the action.

\subsection{\label{secIIsiii}Plebanski general relativity}

The second example of the constrained $BF$ theory is the Plebanski action for general relativity \cite{plebanski1977,BFgravity2016}. Using the Lorentz group $SO(3,1)$ as a gauge group, one constructs a constrained $BF$ action as
\begin{equation}
    S =\int_{\cM_4} B_{ab} \wedge R^{ab} + \phi_{abcd} B^{ab} \wedge B^{cd}\,.
\end{equation}
Here $R^{ab}$ is the curvature $2$-form for the spin connection $\omega^{ab}$, $B_{ab}$ is the usual Lagrange multiplier $2$-form, while $\phi_{abcd}$ is the additional Lagrange multiplier $0$-form multiplying the term $B^{ab}\wedge B^{cd}$ to form a simplicity constraint. It can be shown that the variation of this action with respect to $B_{ab}$, $\omega^{ab}$ and $\phi_{abcd}$ gives rise to the equations of motion of vacuum general relativity. However, in this model the tetrad fields appear only as a solution of the simplicity constraint equation of motion $B^{ab}\wedge B^{cd} = 0$. Therefore, being intrinsically on-shell objects, the tetrad fields are not present in the action itself and cannot be quantized. This renders the Plebanski model unsuitable for coupling of matter fields to gravity \cite{RovelliSpinfoamFermions,MikovicVojinovic2012,VojinovicCDT2016}. Nevertheless, regarded as a model for pure gravity, the Plebanski model has been successfully quantized in the context of spinfoam models \cite{EPRL,FK,RovelliBook,RovelliVidottoBook}.

\section{\label{secIII}$2BF$ theory}

In this section we perform the first step of the {\em categorical ladder}, generalizing the algebraic notion of a group to the notion of a $2$-group. This leads to the generalization of the $BF$ theory to the $2BF$ theory, also sometimes called $BFCG$ theory \cite{BaezHuerta2011, GirelliPfeifferPopescu2008, FariaMartinsMikovic2011, crane2003}.

\subsection{\label{secIIIsi}Pure $2BF$ theory}

In order to circumvent the issue of tetrad fields not being present in the Plebanski action, in the context of higher category theory \cite{BaezHuerta2011} a recent promising approach has been developed \cite{MikovicVojinovic2012,MikovicStrongWeak,MikovicOliveira2014,MOV2016,VojinovicCDT2016,MOV2019}. As an essential ingredient, let us first give a short review of the $2$-group formalism.

Within the framework of category theory, the group as an algebraic structure can be understood as a category with only one object and invertible morphisms \cite{BaezHuerta2011}. Additionally, the notion of a category can be generalized to the so-called {\em higher categories}, which have not only objects and morphisms, but also $2$-morphisms (morphisms between morphisms), and so on. This process of generalization is called the {\em categorical ladder}. Using this process, one can introduce the notion of a {\em $2$-group} as a $2$-category consisting of only one object, where all the morphisms and all $2$-morphisms are invertible. It has been shown that every strict $2$-group is equivalent to a {\em crossed module} $(H \stackrel{\del}{\to}G \,, \rhd)$, see \cite{RadenkovicVojinovic2019} for detailed definitions. Here $G$ and $H$ are groups, $\del$ is a homomorphism from $H$ to $G$, while $\rhd:G\times H \to H$ is an action of $G$ on $H$.

Similarly to the case of an ordinary Lie group $G$ which has a naturally associated notion of a connection $\alpha$, giving rise to a $BF$ theory, the $2$-group structure has a naturally associated notion of a $2$-connection $(\alpha\,,\beta)$, described by the usual $\mathfrak{g}$-valued $1$-form $\alpha \in \cA^1(\cM_4\,,\mathfrak{g})$ and an $\mathfrak{h}$-valued $2$-form $\beta \in \cA^2(\cM_4\,,\mathfrak{h})$, where $\mathfrak{h}$ is a Lie algebra of the Lie group $H$. The $2$-connection gives rise to the so-called {\em fake $2$-curvature} $(\cF,\cG)$, given as
\begin{equation}\label{eq:krivine}
    \cF=\D \alpha+ \alpha \wedge \alpha - \partial\beta\,, \quad \quad \cG= d\beta+\alpha\wedge^\rhd \beta\,.
\end{equation}
Here $\alpha \wedge^\rhd \beta$ means that $\alpha$ and $\beta$ are multiplied as forms using $\wedge$, and simultaneously multiplied as algebra elements using $\rhd$, see \cite{RadenkovicVojinovic2019}. The curvature pair $(\cF,\cG)$ is called ``fake'' because of the presence of the additional term $\partial\beta$ in the definition of $\cF$ \cite{BaezHuerta2011}.

Using the structure of a $2$-group, or equivalently the crossed module, one can generalize the $BF$ action to the so-called $2BF$ action, defined as follows \cite{GirelliPfeifferPopescu2008,FariaMartinsMikovic2011}:
\begin{equation}\label{eq:bfcg}
S_{2BF} =\int_{\cM_4} \langle B \wedge \cF \rangle_{\mathfrak{ g}} +  \langle C \wedge \cG \rangle_{\mathfrak{h}} \,.
\end{equation}
Here the $2$-form $B \in \cA^2(\cM_4\,, \mathfrak{g})$ and the $1$-form $C\in \cA^1(\cM_4\,,\mathfrak{h})$ are Lagrange multipliers. Also, $\langle \_\,,\_\rangle{}_{\mathfrak{g}}$ and $\langle \_\,,\_\rangle{}_{\mathfrak{h}}$ denote the $G$-invariant bilinear symmetric nondegenerate forms for the algebras $\mathfrak{g}$ and $\mathfrak{h}$, respectively. As a consequence of the axiomatic structure of a crossed module (see \cite{RadenkovicVojinovic2019}), the bilinear form $\langle \_\,,\_\rangle{}_{\mathfrak{h}}$ is $H$-invariant as well. See \cite{GirelliPfeifferPopescu2008,FariaMartinsMikovic2011} for review and references.

Similarly to the $BF$ action, the $2BF$ action is also topological, which can be seen from equations of motion. Varying with respect to $B^{\alpha}$ and $C^a$ one obtains
\begin{equation}\label{eq:jedn2bf}
    \cF^{\alpha}=0\,, \quad \quad \cG^a=0\,,
\end{equation}
where indices $a$ count the generators of the group $H$. Varying with respect to $\alpha^{\alpha}$ and $\beta^a$ one obtains the equations for the multipliers,
\begin{gather}
\D B_\alpha+ {f_{\alpha \beta}}^\gamma B_\gamma \wedge  \alpha^\beta - {\rhd_{\alpha a}}^b C_b\wedge \beta^a =0\,, \\ 
\D C_a - {\partial_a}^\alpha B_\alpha + {\rhd_{\alpha a}}^b C_b \wedge \alpha^\alpha =0\,.
\end{gather}
We can again see that the equations of motion are only first-order and have only very simple solutions (note that this is not a sufficient argument for the absence of local propagating degrees of freedom --- a counterexample is the Dirac equation, being a first-order partial differential equation which {\em does} have propagating degrees of freedom). One can additionally use the Hamiltonian analysis to rigorously demonstrate that there are no local propagating degrees of freedom \cite{MikovicOliveira2014,MOV2016}. Thus the $2BF$ theory is also topological.

\subsection{\label{secIIIsii}General relativity}

An important example of a crossed module structure is a vector space $V$ equipped with an isometry group $O$. Namely, $V$ can be regarded as an Abelian Lie group with addition as a group operation, so that a representation of $O$ on $V$ is an action $\rhd$ of $O$ on the group $V$, giving rise to the crossed module $(V \stackrel{\del}{\to}O \,, \rhd)$, where the homomorphism $\del$ is chosen to be trivial (it maps every element of $V$ into a unit of $O$).

We can employ this construction to introduce the {\em Poincar\'e $2$-group}. One constructs a crossed module by choosing
\begin{equation}
G=SO(3,1)\,, \qquad H=\realni^4\,.
\end{equation}
The map $\partial$ is trivial, while $\rhd$ is a natural action of $SO(3,1)$ on $\realni^4$, defined by the equation
\begin{equation} \label{VektorskaRjaLorentza}
M_{ab}\rhd P_c = \eta_{[bc} P_{a]}\,,
\end{equation}
where $M_{ab}$ and $P_a$ are the generators of groups $SO(3,1)$ and $\realni^4$, respectively. The action $\rhd$ of $SO(3,1)$ on itself is given via conjugation. At the level of the algebra, conjugation reduces to the action via the adjoint representation, so that
\begin{equation}
M_{ab} \rhd M_{cd} = \komut{M_{ab}}{M_{cd}} \equiv  \eta_{ad} M_{bc} - \eta_{ac} M_{bd} + \eta_{bc} M_{ad} - \eta_{bd} M_{ac}\,.
\end{equation}
The $2$-connection $(\alpha, \beta)$ is given by the algebra-valued differential forms
\begin{equation}
\alpha=\omega^{ab}M_{ab}\,, \qquad \beta = \beta^a P_a\,,
\end{equation}
where $\omega^{ab}$ is called the spin connection. The corresponding $2$-curvature in this case is given by
\begin{equation}  \label{eq:krivinezapoenc}
  \begin{array}{lclcl}
 {\cal F} & = & (\mathrm{d} \omega^{ab} + {\omega^a}_c\wedge\omega^{cb} )M_{ab} & \equiv & R^{ab}M_{ab} \,, \vphantom{\ds\int} \\
 {\cal G} & = & (\mathrm{d}\beta^a + {\omega^a}_b \wedge \beta^b)P_a & \equiv & \nabla\beta^a P_a \equiv G^a P_a\,, \vphantom{\ds\int} \\
\end{array}
\end{equation}
Note that, since $\del$ is trivial, the fake curvature is the same as ordinary curvature. Introducing the bilinear forms
\begin{equation}
\killing{M_{ab}}{M_{cd}}_{\mathfrak{g}} = \eta_{a[c} \eta_{bd]} \,, \qquad \killing{P_a}{P_b}_{\mathfrak{h}} = \eta_{ab}\,,
\end{equation}
one can show that $1$-forms $C^a$ transform in the same way as the tetrad $1$-forms $e^a$ under the Lorentz transformations and diffeomorphisms, so the fields $C^a$ can be identified with the tetrads. Then one can rewrite the pure $2BF$ action (\ref{eq:bfcg}) for the Poincar\'e $2$-group as
\begin{equation}\label{eq:GravityTopoloski}
S_{2BF} = \int_{\cM_4} B^{ab}\wedge R_{ab} + e_a \wedge \nabla\beta^a \,. 
\end{equation}
Note that the above step of recognizing that $C^a \equiv e^a$ was crucial, since we now see that the tetrad fields are explicitly present in the $2BF$ action for the Poincar\'e $2$-group.

In order to promote (\ref{eq:GravityTopoloski}) to an action for general relativity, we add a convenient simplicity constraint term:
\begin{equation}\label{eq:GravityVeza}
  S = \int_{\cM_4} B^{ab}\wedge R_{ab} + e_a \wedge \nabla\beta^a - \lambda_{ab} \wedge \Big( B^{ab}-\frac{1}{16\pi l_p^2}\varepsilon^{abcd} e_c \wedge e_d \Big) \,.
\end{equation}
Here $\lambda_{ab}$ is a Lagrange multiplier $2$-form associated to the simplicity constraint term, and $l_p$ is the Planck length. Note that the term ``simplicity constraint'' derives its name from the fact that the constraint imposes the property of {\em simplicity} on $B^{ab}$ --- a $2$-form is said to be {\em simple} if it can be written as an exterior product of two $1$-forms.

Varying the action (\ref{eq:GravityVeza}) with respect to $B_{ab}$, $e_{a}$, $\omega_{ab}$, $\beta_{a}$ and $\lambda_{ab}$, we obtain the following equations of motion:
\begin{gather}  
\label{eq:g3a}
R_{ab} - \lambda_{ab} = 0\,,\vphantom{\ds\int}\\
\label{eq:g4a}
\nabla \beta_a + \frac{1}{8\pi l_p^2} \varepsilon_{abcd} \lambda^{bc} \wedge e^d=0\,,\vphantom{\ds\int}\\
\label{eq:g5a}
\nabla B_{ab} - e_{[a} \wedge \beta_{b]} = 0\,,\vphantom{\ds\int}\\
\label{eq:g6a}
\nabla e_a = 0\,,\vphantom{\ds\int}\\
\label{eq:g7a}
B^{ab}-\frac{1}{16\pi l_p^2}\varepsilon^{abcd} e_c \wedge e_d=0\,.\vphantom{\ds\int}
\end{gather}
Given this system of equations, all fields can be algebraically determined in terms of the tetrads $e^a{}_{\mu}$, as follows. From the equations (\ref{eq:g6a}) and (\ref{eq:g7a}) we obtain that $\nabla B^{ab} = 0$, from which it follows, using the equation (\ref{eq:g5a}), that $e_{[a} \wedge \beta_{b]} = 0$. Assuming that the tetrads are nondegenerate, $e\equiv \det (e^a{}_{\mu}) \neq 0$, it can be shown that this is equivalent to $\beta^{a}=0$ \cite{MikovicVojinovic2012}. Therefore, from the equations (\ref{eq:g3a}), (\ref{eq:g5a}), (\ref{eq:g6a}) and (\ref{eq:g7a}) we obtain
\begin{equation}
    \lambda^{ab}{}_{\mu\nu}=R^{ab}{}_{\mu\nu}\,,  \quad
    \beta^a{}_{\mu\nu}=0\,, \quad
    B_{ab}{}_{\mu \nu}=\frac{1}{8\pi l_p^2}\varepsilon_{abcd} e^c{}_\mu e^d{}_\nu\,, \quad
    \omega^{ab}{}_\mu =\triangle^{ab}{}_\mu\,.
\end{equation}
Here the Ricci rotation coefficients are defined as
\begin{equation}
    \triangle^{ab}{}_\mu\equiv \frac{1}{2}(c^{abc}-c^{cab}+c^{bca})e_{c\mu}\,, 
\end{equation}
where
\begin{equation}
    c^{abc}= e^{\mu}{}_b e^{\nu}{}_c \left( \partial_\mu e^a{}_\nu - \partial_\nu e^a{}_\mu \right)\,.
\end{equation}
The last equation establishes that the spin connection $1$-form $\omega^{ab}$ is expressed as a function of the tetrads, which then implies the same for the curvature $2$-form $R^{ab}$. Finally, the remaining equation (\ref{eq:g4a}) then reduces to
\begin{equation}
 \varepsilon_{abcd} R^{bc} \wedge e^d=0\,,
\end{equation}
which is nothing but the vacuum Einstein field equation,
$$
R_{\mu\nu}-\frac{1}{2}g_{\mu\nu}R = 0\,.
$$
Therefore, the action (\ref{eq:GravityVeza}) is classically equivalent to general relativity.

\subsection{\label{secIIIsiii}Einstein-Yang-Mills theory}

As we have already mentioned above, the main advantage of the action (\ref{eq:GravityVeza}) over the Plebanski model lies in the fact that the tetrad fields are explicitly present in the topological sector of the action. This allows one to couple matter fields in a straightforward way \cite{MikovicVojinovic2012}. However, one can do even more \cite{RadenkovicVojinovic2019}, and couple the $SU(N)$ Yang-Mills fields to gravity within a unified framework of $2$-group formalism.

Namely, we can modify the Poincar\'e $2$-group structure to include the $SU(N)$ gauge group, as follows. We choose the two Lie groups as
\begin{equation}
G=SO(3,1)\times SU(N)\,, \qquad H=\mathbb{R}^4\,,
\end{equation}
and we define the action $\rhd$ of the group $G$ in the following fashion. As in the case of the Poincar\'e $2$-group, it acts on itself via conjugation. Next, it acts on $H$ such that the $SO(3,1)$ subgroup acts on $\realni^4$ via the vector representation (\ref{VektorskaRjaLorentza}), while the action of the $SU(N)$ subgroup is trivial,
\begin{equation} \label{DelovanjeTauNaPeove}
\tau_I \rhd P_a = 0\,,
\end{equation}
where $\tau_I$ are the $SU(N)$ generators. The map $\partial$ also remains trivial, as before. The form of the $2$-connection $(\alpha,\beta)$ now reflects the structure of the group $G$,
\begin{equation}
\alpha = \omega^{ab}M_{ab} + A^I \tau_I\,, \qquad \beta = \beta^a P_a\,,
\end{equation}
where $A^I$ is the gauge connection $1$-form. Next, the curvature for $\alpha$ then becomes
\begin{equation}
{\cal F} = R^{ab}M_{ab} + F^I \tau_I\,, \qquad F^I \equiv \rmd A^I + f_{JK}{}^I A^J \wedge A^K\,.
\end{equation}
The curvature for $\beta$ remains the same as before, because of (\ref{DelovanjeTauNaPeove}). Finally, the product structure of the group $G$ implies that its Killing form $\killing{\_}{\_}_{\mathfrak{g}}$ reduces to the Killing forms for the $SO(3,1)$ and $SU(N)$, along with the identity $\killing{M_{ab}}{\tau_I}_{\mathfrak{g}}=0$.

Given a crossed module defined in this way, its corresponding pure $2BF$ action (\ref{eq:bfcg}) becomes
\begin{equation} \label{eq:bfcggauge}
    S_{2BF}=\int_{\cM_4} B^{ab}\wedge R_{ab} + B^I \wedge F_I + e_a \wedge \nabla \beta^a\,,
\end{equation}
where $B^I \in \cA^2(\cM_4\,,\mathfrak{su}(N))$ is the new Lagrange multiplier. The action (\ref{eq:bfcggauge}) is topological, and again we add appropriate simplicity constraint terms, in order to transform it into action with nontrivial dynamics. The constraint giving rise to gravity is the same as in (\ref{eq:GravityVeza}), while the constraint for the gauge fields is given as in the action (\ref{eq:bfgauge}) with the substitution $\delta^a \to e^a$. Putting everything together, we obtain:
\begin{equation} \label{eq:YMplusGravity}
\begin{aligned}
S=&\int_{\cM_4} B^{ab}\wedge R_{ab} + B^I \wedge F_I + e_a \wedge \nabla \beta^a \\
  &- \lambda_{ab} \wedge \Big(B^{ab}-\frac{1}{16\pi l_p^2}\varepsilon^{abcd} e_c \wedge e_d\Big) + \lambda^I\wedge \Big(B_I-\frac{12}{g}{M_{ab}}_Ie^a\wedge e^b\Big) \\
  & + {\zeta^{ab}}{}^I \Big( {M_{ab}}{}_I\varepsilon_{cdef}e^c\wedge e^d \wedge e^e \wedge e^f- g_{IJ}F^J \wedge e_a \wedge e_b \Big) \,. \\
\end{aligned}
\end{equation}
It is crucial to note that the Yang-Mills simplicity constraints in (\ref{eq:YMplusGravity}) are obtained from the Yang-Mills action (\ref{eq:bfgauge}) by substituting the nondynamical background field $\delta^a$ from (\ref{eq:bfgauge}) with a dynamical field $e^a$. The relationship between these fields has already been hinted at in the equation (\ref{eq:flatspacetimemetric}), which describes the connection between $\delta^a$ and the flat spacetime metric $\eta_{\mu\nu}$. Once promoted to $e^a$, this field becomes dynamical due to the presence of gravitational terms, while the equation (\ref{eq:flatspacetimemetric}) becomes the usual relation between the tetrad and the metric,
\begin{equation}
g_{\mu\nu} = \eta_{ab} e^a{}_{\mu} e^b{}_{\nu}\,,
\end{equation}
further confirming the identification $C^a = e^a$. Moreover, the total action (\ref{eq:YMplusGravity}) now becomes background independent, as expected in general relativity. All this is a consequence of the fact that the tetrad field is explicitly present in the topological sector of the action (\ref{eq:GravityVeza}), and represents a clear improvement over the Plebanski model.

Taking the variations of the action (\ref{eq:YMplusGravity}) with respect to the variables $B_{ab}$, $\omega_{ab}$, $\beta_a$, $\lambda_{ab}$, ${\zeta^{ab}}{}^I$, ${M_{ab}}{}_I$, $B_I$, $\lambda^I$, $A^I$, and $e^a$, we obtain equations of motion. Similarly as before, all variables can be algebraically expressed as functions of $A^I$ and $e^a$ and their derivatives:
\begin{equation}\label{eq:012}
    \begin{array}{c}
        \lambda_{ab}{}_{\mu\nu}=R_{ab}{}_{\mu\nu}\,,\qquad  \beta_a{}_{\mu \nu}=0\,,\qquad \omega_{ab}{}_\mu=\triangle_{ab}{}_\mu\,, \qquad  \lambda{}_{ab}{}_I=F_{ab}{}_I\,, \vphantom{\ds\int} \\
 \ds B_{\mu\nu}{}_I=-\frac{e}{2g}{\varepsilon_{\mu\nu\rho\sigma}} {F^{\rho \sigma}}_I\,, \qquad B_{ab}{}_{\mu \nu}=\frac{1}{8\pi l_p^2}\varepsilon_{abcd} e^c{}_\mu e^d{}_\nu\,,\\
 \ds M_{ab}{}_I=-\frac{1}{4eg}\varepsilon^{\mu \nu \rho \sigma}F_{\mu \nu}{}^Ie^a{}_\rho e^b{}_\sigma\,, \qquad  {\zeta^{ab}}{}^I=\frac{1}{4eg}\varepsilon^{\mu \nu \rho \sigma}F_{\mu \nu}{}^Ie^a{}_\rho e^b{}_\sigma\,. \\
    \end{array}
\end{equation}
In addition, we obtain two differential equations --- An equation for $A^I$,
\begin{equation}
   \nabla_\rho F^{I \rho \mu}\equiv  \partial_\rho F^{I\rho\mu} + \itGamma^{\rho}{}_{\lambda\rho} F^{I\lambda\mu} + f_{JK}{}^I A^J{}_\rho F^{K\rho\mu} =0\,,\label{eq:020}
\end{equation}
where $\itGamma^{\lambda}{}_{\mu\nu}$ is the standard Levi-Civita connection, and an equation for $e^a$,
\begin{equation}
  R^{\mu \nu}-\frac{1}{2}g^{\mu \nu}R=8\pi l_p^2 \; T^{\mu \nu}\,,
\end{equation}
where
\begin{equation}
  T^{\mu \nu} \equiv -\frac{1}{4g}\left(F_{\rho \sigma}{}^IF^{\rho \sigma}{}_Ig^{\mu \nu}+4F^{\mu \rho}{}_I{F_\rho}^{\nu}{}^I \right)\,.\label{eq:019}
\end{equation}

In this way, we see that both gravity and gauge fields can be successfully represented within a unified framework of higher gauge theory, based on a $2$-group structure. A generalization from $SU(N)$ Yang-Mills case to more complicated cases such as $SU(3)\times SU(2) \times U(1)$ is completely straightforward.

\section{\label{secIV}$3BF$ theory}

While the structure of a $2$-group can successfully describe both gravitational and gauge fields, unfortunately it cannot accommodate other matter fields, such as scalars or fermions. In order to remedy this drawback, we make one further step in the categorical ladder, passing from the notion of a $2$-group to the notion of a $3$-group. As it turns out, the $3$-group structure is excellent for the description of all fields that are present in the Standard Model, coupled to gravity. Moreover, a $3$-group contains one more gauge group, which is novel and corresponds to the choice of the scalar and fermion fields present in the theory. This is an unexpected and beautiful result, not present in ordinary gauge theory.

As before, we will begin by introducing the notion of a $3$-group, and constructing the corresponding $3BF$ action. Afterwards, we will modify this action by adding appropriate simplicity constraints, giving rise to theories with expected nontrivial dynamics. Along the way, we shall see that scalar and fermion fields are being treated pretty much on an equal footing with gravity and gauge fields.

\subsection{\label{secIVsi}Pure $3BF$ theory}

Similarly to the concepts of a group and a $2$-group, one can introduce the notion of a $3$-group in the framework of higher category theory, as a $3$-category with only one object where all the morphisms, $2$-morphisms and $3$-morphisms are invertible. Also, in the same way as a $2$-group is equivalent to a crossed module, it was proved that a strict $3$-group is equivalent to a {\em $2$-crossed module} \cite{martins2011}.

A Lie $2$-crossed module, denoted as $(L\stackrel{\delta}{\to} H \stackrel{\partial}{\to}G\,, \rhd\,, \{\_\,,\_\})$, is an algebraic structure specified by three Lie groups $G$, $H$ and $L$, together with the homomorphisms $\delta$ and $\del$, an action $\rhd$ of the group $G$ on all three groups, and a $G$-equivariant map
\begin{displaymath}
\{ \_ \,, \_ \} : H\times H \to L\,.
\end{displaymath}
called the Peiffer lifting. The maps $\del$, $\delta$, $\rhd$ and the Peiffer lifting satisfy certain axioms, so that the resulting structure is equivalent to a $3$-group \cite{RadenkovicVojinovic2019}.

Like in the cases of $BF$ and $2BF$ actions, we can introduce a gauge invariant topological $3BF$ action over the manifold $\mathcal{M}_4$ for a given $2$-crossed module $(L\stackrel{\delta}{\to} H \stackrel{\partial}{\to}G\,, \rhd\,, \{\_\,,\_\})$. Denoting $\mathfrak{g}$, $\mathfrak{h}$ and $\mathfrak{l}$ as Lie algebras corresponding to the groups $G$, $H$ and $L$, respectively, one can introduce a $3$-connection $(\alpha, \beta,\gamma)$ given by the algebra-valued differential forms $\alpha \in \cA^1(\cM_4\,,\mathfrak{g})$, $\beta \in \cA^2(\cM_4\,,\mathfrak{h})$ and $\gamma \in \cA^3(\cM_4\,,\mathfrak{l})$. The corresponding fake $3$-curvature $(\cal F\,, G\,, H)$ is then defined as
\begin{equation}\label{eq:3krivine}
\begin{array}{c}
\ds  \cF = \D \alpha+\alpha \wedge \alpha - \partial \beta \,, \quad \quad \cG = \D \beta + \alpha \wedge^\rhd \beta - \delta \gamma\,, \vphantom{\ds\int} \\
\ds  \cH = \D \gamma + \alpha\wedge^\rhd \gamma + \{\beta \wedge \beta\} \,,
\end{array}
\end{equation}
see \cite{martins2011, Wang2014} for details. Note that $\gamma$ is a $3$-form, while its corresponding field strength $\cH$ is a $4$-form, necessitating that the spacetime manifold be at least $4$-dimensional. Then, a $3BF$ action is defined as
\begin{equation}\label{eq:bfcgdh}
S_{3BF} =\int_{\mathcal{M}_4} \langle B \wedge  {\cal F} \rangle_{\mathfrak{ g}} +  \langle C \wedge  {\cal G} \rangle_{\mathfrak{h}} + \langle D \wedge {\cal H} \rangle_{\mathfrak{l}} \,,
\end{equation}
where $B \in \cA^2(\cM_4,\mathfrak{g})$, $C \in \cA^1(\cM_4,\mathfrak{h})$ and $D \in \cA^0(\cM_4,\mathfrak{l})$ are Lagrange multipliers. Note that in precisely $4$ spacetime dimensions the Lagrange multiplier $D$ corresponding to $\cH$ is a $0$-form, i.e. a scalar function. The functionals $\killing{\_}{\_}_{\mathfrak{g}}$, $\killing{\_}{\_}_{\mathfrak{h}}$ and $\killing{\_}{\_}_{\mathfrak{l}}$ are $G$-invariant bilinear symmetric nondegenerate forms on $\mathfrak{g}$,  $\mathfrak{h}$ and $\mathfrak{l}$, respectively. Under certain conditions, the forms $\killing{\_}{\_}_{\mathfrak{h}}$ and $\killing{\_}{\_}_{\mathfrak{l}}$ are also $H$-invariant and $L$-invariant.

One can see that varying the action with respect to the variables $B^{\alpha}$, $C^a$ and $D^A$ (where indices $A$ count the generators of the group $L$), one obtains the equations of motion 
\begin{equation}\label{eq:jed3bf}
\mathcal{F}^{\alpha} = 0 \,, \quad \quad \mathcal{G}^a = 0\,, \quad \quad \mathcal{H}^A = 0 \,,
\end{equation}
while varying with respect to $\alpha^{\alpha}$, $\beta^a$, $\gamma^A$ one obtains
\begin{gather}
\D B_\alpha+ {f_{\alpha \beta}}^\gamma B_\gamma \wedge  \alpha^\beta - {\rhd_{\alpha a}}^b C_b\wedge \beta^a+\rhd_{\alpha B}{}^A D_A\wedge\gamma^B=0\,, \\ 
\D C_a - {\partial_a}^\alpha B_\alpha + {\rhd_{\alpha a}}^b C_b \wedge \alpha^\alpha + 2X_{\{ab\}}{}^AD_A\wedge \beta^b=0\,,\\
\D D_A - \rhd_{ \alpha A}{}^B D_B\wedge \alpha^\alpha+\delta_A {}^a C_a=0\,.
\end{gather}

\subsection{\label{secIVsii}Klein-Gordon theory}

Now we proceed to demonstrate that one can use the $3$-group structure and the corresponding $3BF$ theory to describe the Klein-Gordon field coupled to general relativity. We begin by specifying a $2$-crossed module, which is used to construct the topological $3BF$ theory, and then we impose appropriate simplicity constraints to obtain the desired equations of motion.

We specify a $2$-crossed module $(L\stackrel{\delta}{\to} H \stackrel{\partial}{\to}G\,, \rhd\,, \{\_\,,\_\})$, as follows. The groups are given as
\begin{equation}
G=SO(3,1)\,, \quad \quad H=\mathbb{R}^4\,, \quad \quad L=\mathbb{R}\,.
\end{equation}
The group $G$ acts on itself via conjugation, on $H$ via the vector representation, and on $L$ via the trivial representation. This specifies the definition of the action $\rhd$. The map $\partial$ is chosen to be trivial, as before. The map $\delta$ is also trivial, that is, every element of $L$ is mapped to the identity element of $H$. Finally, the Peiffer lifting is trivial as well, mapping every ordered pair of elements in $H$ to an identity element in $L$. This specifies one concrete $2$-crossed module which, as we shall see below, corresponds to gravity and one real scalar field.

Given this choice of a $2$-crossed module, the $3$-connection $(\alpha\,,\beta\,,\gamma)$ takes the form
\begin{equation}
 \alpha = \omega^{ab}M_{ab}\,, \quad \quad \beta=\beta^a P_a\,, \quad \quad \gamma = \gamma \mathbb{I}\,,
\end{equation}
where $\mathbb{I}$ is the sole generator of the Lie group $\realni$. Since the homomorphisms $\del$ and $\delta$ are trivial, as well as the Peiffer lifting, the fake $3$-curvature (\ref{eq:3krivine}) reduces to the ordinary $3$-curvature,
\begin{equation}
\cF = R^{ab} M_{ab}\,, \quad \quad \cG= \nabla \beta^a P_a\,, \quad \quad \cH= \D \gamma\,,
\end{equation}
where we used the fact that $G$ acts trivially on $L$, that is, $M_{ab}\rhd \mathbb{I}=0$. This means that the $3$-form $\gamma$ transforms as a scalar with respect to Lorentz symmetry. Consequently, its Lagrange multiplier $D$ also transforms as a scalar, since it also belongs to the algebra $\mathfrak{l}$. Since $D$ is also a $0$-form, it transforms as a scalar with respect to diffeomorphisms as well. In other words, $D$ completely behaves as a real scalar field, so we relabel it into more traditional notation, $D \equiv \phi$, and write the pure $3BF$ action (\ref{eq:bfcgdh}) as:
\begin{equation}\label{eq:Scalartopoloski}
    S_{3BF}=\int_{\cM_4} B^{ab}\wedge R_{ab} + e_a\wedge \nabla \beta^a + \phi \, \D \gamma\,,
\end{equation}
where the bilinear form for $L$ is $\killing{\mathbb{I}}{\mathbb{I}}_{\mathfrak{l}} = 1$.

The existence of a scalar field in the $3BF$ action is a crucial property of a $3$-group in a $4$-dimensional spacetime, just like identifying the Lagrange multiplier $C^a$ with a tetrad field $e^a$ was a crucial property of the $2BF$ action and the Poincar\'e $2$-group. We can also see that the choice of the third gauge group, $L$, dictates the number and the structure of the matter fields present in the action. In this case, $L=\realni$ implies that we have only one real scalar field, corresponding to a single generator $\mathbb{I}$ of $\realni$. The trivial nature of the action $\rhd$ of $SO(3,1)$ on $\realni$ implies that $\phi$ transforms as a scalar field. Finally, the scalar field appears in the topological sector of the action, making the quantization procedure feasible.

As in the case of $BF$ and $2BF$ theories, we need to add appropriate simplicity constraints to the action (\ref{eq:Scalartopoloski}). In order to obtain the Klein-Gordon field $\phi$ of mass $m$ coupled to gravity in the standard way, the action takes the form:
\begin{equation}\label{eq:scalar}
\begin{aligned}
 S =\int_{\cM_4} & B^{ab}\wedge R_{ab} + e_a\wedge \nabla \beta^a + \phi \, \D \gamma \vphantom{\ds\int} \\
 &- \lambda_{ab} \wedge \Big(B^{ab}-\frac{1}{16\pi l_p^2}\varepsilon^{abcd} e_c \wedge e_d\Big)\vphantom{\ds\int} \\
 &+ {\lambda}\wedge \Big(\gamma - \frac{1}{2} H_{abc} e^a \wedge e^b \wedge e^c\Big) \\
 &+\Lambda^{ab}\wedge \Big( H_{abc}\varepsilon^{cdef}e_d\wedge e_e \wedge e_f- \D \phi \wedge e_a \wedge e_b\Big) \vphantom{\ds\int} \\
 &-\frac{1}{2\cdot 4!} m^2\phi^2 \varepsilon_{abcd}e^a\wedge e^b \wedge e^c \wedge e^d\,.\vphantom{\ds\int}
\end{aligned}
\end{equation}
The first row is the topological sector (\ref{eq:Scalartopoloski}), the second row is the familiar simplicity constraint for gravity from the action (\ref{eq:GravityVeza}), the third and fourth rows contain the new simplicity constraints featuring the Lagrange multiplier $1$-forms $\lambda$ and $\Lambda^{ab}$ and the $0$-form $H_{abc}$, while the fifth row is the mass term for the scalar field.

The variation of (\ref{eq:scalar}) with respect to the variables $B_{ab}$, $\omega_{ab}$, $\beta_a$, $\lambda_{ab}$, $\Lambda_{ab}$, $\gamma$, ${\lambda}$, $H_{abc}$, $\phi$ and $e^a$ gives us the equations of motion. As before, all variables can be algebraically expressed in terms of the tetrads $e^a$ and the scalar field $\phi$:
\begin{equation}
\begin{array}{c}\label{eq:sys1}
\ds \lambda_{ab}{}_{\mu\nu}=R_{ab}{}_{\mu\nu}\,, \qquad    \omega^{ab}{}_\mu=\triangle^{ab}{}_\mu\,, \qquad  \gamma_{\mu\nu\rho}=-\frac{e}{2}\varepsilon_{\mu\nu\rho\sigma}\partial^\sigma\phi\,, \vphantom{\ds\int} \\
\ds \beta^a{}_{\mu\nu}=0\,, \qquad \Lambda^{ab}{}_{\mu}=\frac{1}{12e}g_{\mu\lambda}\varepsilon^{\lambda\nu\rho\sigma}\partial_\nu\phi e{}^a{}_{\rho} e{}^b{}_\sigma\,,\vphantom{\ds\int} \qquad  \lambda_{\mu}=\partial_{\mu}\phi\,,\vphantom{\ds\int} \\
\ds H^{abc}=\frac{1}{6e} \varepsilon^{\mu\nu\rho\sigma}\partial_\mu\phi e^a{}_\nu e^b{}_\rho e^c{}_\sigma\,, \qquad  B_{ab}{}_{\mu\nu}=\frac{1}{8\pi l_p^2}\varepsilon_{abcd}e^c{}_\mu e^d{}_\nu\,. \\
\end{array}
\end{equation}
The equations of motion for $e^a$ and $\phi$, however, are differential equations. The equation for the scalar field becomes the covariant Klein-Gordon equation,
\begin{equation}
\left(\nabla_\mu\nabla^\mu -m^2\right)\phi=0\,,
\end{equation}
while the equation for the tetrads is
\begin{equation}\label{eq:scalareomfore}
  {R}^{\mu\nu}-\frac{1}{2}g^{\mu\nu} R=8\pi l_p^2\; T^{\mu\nu}\,,
\end{equation}
where
\begin{equation}
T^{\mu\nu}\equiv\partial^\mu \phi \partial^\nu \phi -\frac{1}{2}g^{\mu\nu} \left(\partial_\rho \phi \partial^\rho \phi+m^2\phi^2 \right)
\end{equation}
is the stress-energy tensor for a single real scalar field.

\subsection{\label{secIVsiii}Einstein-Cartan-Dirac theory}

In order to describe the Dirac field coupled to Einstein-Cartan gravity, we follow the same procedure as for the case of the scalar field, but now we choose the $2$-crossed module $(L\stackrel{\delta}{\to} H \stackrel{\partial}{\to}G\,, \rhd\,, \{\_\,,\_\})$ in a different way, as follows. The groups are:
\begin{equation}
G=SO(3,1)\,, \quad \quad H=\mathbb{R}^4\,, \quad \quad L=\mathbb{R}^8(\grasmanovi) \,,
\end{equation}
where $\grasmanovi$ is the algebra of complex Grassmann numbers. The maps $\partial$,  $\delta$ and the Peiffer lifting are trivial, as before. The action of the group $G$ on itself is given via conjugation, on $H$ via vector representation, and on $L$ via spinor representation, in the following way. Denoting the $8$ generators of the Lie group $\mathbb{R}^8(\grasmanovi)$ as $P_{\alpha}$ and $P^{\alpha}$, where the index $\alpha$ takes the values $1,\dots,4$, the action $\rhd$ of $G$ on $L$ is thus given explicitly as
\begin{equation} \label{eq:actionOfGonLdirac}
M_{ab}\rhd P_{\alpha}=\frac{1}{2}(\sigma_{ab}){}^{\beta}{}_{\alpha} P_{\beta}\,, \qquad M_{ab} \rhd P^{\alpha}=-\frac{1}{2}(\sigma_{ab}){}^{\alpha}{}_{\beta} P^{\beta}\,,
\end{equation}
where $\sigma_{ab}=\frac{1}{4}[\gamma_a,\gamma_b]$, and $\gamma_a$ are the usual Dirac matrices, satisfying the anticommutation rule $\{ \gamma_a\,,\, \gamma_b  \} = -2\eta_{ab}$.

As in the case of the scalar field, the choice of the group $L$ dictates the matter content of the theory, while the action $\rhd$ of $G$ on $L$ specifies its transformation properties.

Let us now proceed to construct the $3BF$ action. The $3$-connection $(\alpha\,,\beta\,,\gamma)$ takes the form
\begin{equation}
\alpha = \omega^{ab}M_{ab}\,, \quad \quad \beta=\beta^a P_a\,, \quad \quad \gamma = \gamma^{\alpha} P_{\alpha}+\bar{\gamma}{}_{\alpha} P^{\alpha}\,,
\end{equation}
while the $3$-curvature $(\cF\,, \cG\,, \cH)$ is given as
\begin{equation}
\begin{aligned}
    \cF = R^{ab} M_{ab}\,, \quad & \quad \cG= \nabla \beta^a P_a\,,\\
    \cH= \Big(\D \gamma^{\alpha}+\frac{1}{2}\omega^{ab}(\sigma_{ab}){}^{\alpha}{}_{\beta}\gamma^{\beta}\Big) P_{\alpha} + & \Big(\D \bar{\gamma}{}_{\alpha}-\frac{1}{2}\omega^{ab}\bar{\gamma}_{\beta}(\sigma_{ab}){}^{\beta}{}_{\alpha}\Big)P^{\alpha} \\
    \equiv (\overset{\rightarrow}{\nabla} \gamma){}^{\alpha} P_{\alpha}+ (\bar{\gamma}\overset{\leftarrow}{\nabla}){}_{\alpha}P^{\alpha}\,, \hphantom{mmm} &
\end{aligned} 
\end{equation}
where we have used (\ref{eq:actionOfGonLdirac}).
The bilinear form $\killing{\_}{\_}_{\mathfrak{l}}$ is defined via its action on the generators:
\begin{equation} \label{eq:DiracKillingForm}
  \begin{array}{c}
\killing{P_{\alpha}}{P_{\beta}}_{\mathfrak{l}} = 0\,, \qquad
\killing{P^{\alpha}}{P^{\beta}}_{\mathfrak{l}} = 0\,, \vphantom{\ds\int}\\
\killing{P_{\alpha}}{P^{\beta}}_{\mathfrak{l}} = - \delta^{\beta}_{\alpha}\,, \qquad
\killing{P^{\alpha}}{P_{\beta}}_{\mathfrak{l}} = \delta_{\beta}^{\alpha}\,. \vphantom{\ds\int} \\
\end{array}
  \end{equation}
Note that the bilinear form defined in this way is antisymmetric, rather than symmetric, when it acts on the generators. The reason for this is the following. For general $A,B\in\mathfrak{l}$, we want the bilinear form to be symmetric. Expanding $A$ and $B$ into components, we can write
\begin{equation}
\killing{A}{B}_{\mathfrak{l}} = A^I B^J g_{IJ}\,, \qquad \killing{B}{A}_{\mathfrak{l}} = B^J A^I g_{JI}\,.
\end{equation}
Since we require the bilinear form to be symmetric, the two expressions must be equal. However, since the coefficients in $\mathfrak{l}$ are Grassmann numbers, we have $A^IB^J= -B^JA^I$, so it follows that $g_{IJ} = - g_{JI}$. Hence the antisymmetry of (\ref{eq:DiracKillingForm}) --- it compensates for the anticommutativity property of the Grassman coefficients, making the bilinear form symmetric for general algebra elements $A,B\in\mathfrak{l}$.

Now we employ the action $\rhd$ of $G$ on $L$ to determine the transformation properties of the Lagrange multiplier $D$ in (\ref{eq:bfcgdh}). Indeed, the choice of the group $L$ dictates that $D$ contains $8$ independent complex Grassmannian matter fields as its components. Moreover, due to the fact that $D$ is a $0$-form and that it transforms according to the spinorial representation of $SO(3,1)$, we can identify its components with the Dirac bispinor fields, and write
\begin{equation}
D = \psi^{\alpha} P_{\alpha} + \bar{\psi}_{\alpha} P^{\alpha}\,.
\end{equation}
This is again an illustration of the fact that information about the structure of the matter sector in the theory is specified by the choice of the group $L$ in the $2$-crossed module, and its transformation properties with respect to the Lorentz group are fixed by the action $\rhd$.

Given all of the above, we write the corresponding pure $3BF$ action as:
\begin{equation}\label{eq:DiracTopoloski}
    S_{3BF}=\int_{\cM_4} B^{ab}\wedge R_{ab} + e_a\wedge \nabla \beta^a + (\bar{\gamma}{\overset{\leftarrow}{\nabla}) {}_{\alpha} \psi^{\alpha} +\bar{\psi}_{\alpha}{({\overset{\rightarrow}{\nabla}}\gamma)}{}^{\alpha}}\,.
\end{equation}
In order to obtain the action that gives us the dynamics of Einstein-Cartan theory of gravity coupled to a Dirac field, we add the following simplicity constraints:
\begin{equation}\label{eq:Dirac}
\begin{aligned}
S= & \int_{\cM_4}  B^{ab}\wedge R_{ab} + e_a\wedge \nabla \beta^a + ( \bar{\gamma} {\overset{\leftarrow}{\nabla}} ) {}_{\alpha} \psi^{\alpha} +\bar{\psi}_{\alpha}{({\overset{\rightarrow}{\nabla}}\gamma)}{}^{\alpha} \\
&- \lambda_{ab} \wedge \Big(B^{ab}\vphantom{\ds\int}-\frac{1}{16\pi l_p^2}\varepsilon^{abcd} e_c \wedge e_d\Big)\vphantom{\ds\int}\\
&-\lambda^{\alpha}\wedge \Big({\bar{\gamma}}_{\alpha}-\frac{i}{6}\varepsilon_{abcd}e^a\wedge e^b \wedge e^c (\bar{\psi}\gamma^d)_{\alpha}\Big)\vphantom{\ds\int}\\
&+\bar{\lambda}_{\alpha}\wedge \Big({\gamma}^{\alpha}+\frac{i}{6}\varepsilon_{abcd}e^a\wedge e^b \wedge e^c (\gamma^d\psi){}^{\alpha}\Big)\vphantom{\ds\int}\\
 & -\frac{1}{12} m \, \bar{\psi}\psi\, \varepsilon_{abcd}e^a\wedge e^b \wedge e^c \wedge e^d +2 \pi i l_p^2 \, \bar{\psi}\gamma_5\gamma^a \psi \, \varepsilon_{abcd}e^b\wedge e^c \wedge \beta^d .\vphantom{\ds\int}
\end{aligned}
\end{equation}
Similarly to the previous case of the scalar field, we recognize the topological sector in the first row, the gravitational simplicity constraint in the second row, while the third and fourth rows contain the new simplicity constraints for the Dirac field, featuring the Lagrange multiplier $1$-forms $\lambda^{\alpha}$ and $\bar{\lambda}_{\alpha}$. The fifth row contains the mass term for the Dirac field, and a term which ensures the correct coupling between the torsion and the spin of the Dirac field. In particular, we want to obtain
\begin{equation} \label{eq:DiracTorzija}
T_a \equiv \nabla e_a = 2\pi l_p^2 s_a\,,
\end{equation}
as one of the equations of motion, where
\begin{equation}\label{eq:torzija}
s_a = i\varepsilon_{abcd} e^b \wedge e^c \bar\psi \gamma_5 \gamma^d \psi
\end{equation}
is the Dirac spin $2$-form. Of course, other alternative coupling choices are possible, but we choose this one since this is the traditional coupling most often discussed in textbooks.

The variation of the action (\ref{eq:Dirac}) with respect to $B_{ab}$, $\lambda^{ab}$, $\bar{\gamma}_{\alpha}$, $\gamma^{\alpha}$, $\lambda^{\alpha}$, ${\bar{\lambda}}_{\alpha}$, $\bar{\psi}{}_{\alpha}$, $\psi^{\alpha}$, $e^a$, $\beta^a$ and $\omega^{ab}$, again gives us equations of motion, which can be algebraically solved for all fields as functions of $e^a$, $\psi$ and $\bar{\psi}$:
\begin{equation}
    \begin{array}{c}\label{eq:d9}
 \ds   B_{ab}{}_{\mu\nu}=\frac{1}{8\pi l_p^2}\varepsilon_{abcd}e^c{}_\mu e^d{}_\nu\,, \qquad  \lambda^ {\alpha}{}_\mu=(\overset{\rightarrow}{\nabla}_\mu\psi)^{\alpha}\,, \qquad   {\bar{\lambda}}_{\alpha}{}_\mu=(\bar{\psi}\overset{\leftarrow}{\nabla}_\mu)_{\alpha}\,, \vphantom{\ds\int}\\
    \bar{\gamma}{}_{\alpha}{}_{\mu\nu\rho}=i\varepsilon_{abcd}e^a{}_\mu e^b{}_\nu  e^c{}_\rho(\bar{\psi}\gamma^d)_{\alpha}\,, \qquad
    {\gamma}^{\alpha}{}_{\mu\nu\rho}=-i\varepsilon_{abcd}e^a{}_\mu e^b{}_\nu  e^c{}_\rho (\gamma^d\psi)^{\alpha}\,, \vphantom{\ds\int} \\
\beta^a{}_{\mu\nu} = 0\,, \qquad \lambda_{ab}{}_{\mu\nu}=R_{ab}{}_{\mu\nu}\,, \qquad \omega^{ab}{}_\mu=\triangle^{ab}{}_\mu+K^{ab}{}_\mu\,. \vphantom{\ds\int} \\
\end{array}
\end{equation}
Here $K^{ab}{}_{\mu}$ is the contorsion tensor, constructed in the standard way from the torsion tensor. In addition, we also obtain
\begin{equation} \label{eq:DiracTorzijaJosJednom}
T_a \equiv \nabla e_a = 2\pi l_p^2 s_a\,,
\end{equation}
which is precisely the desired equation (\ref{eq:DiracTorzija}) for the torsion. Finally, the differential equations of motion for $\psi$ and $\bar{\psi}$ are the standard covariant Dirac equation,
\begin{equation} \label{eq:DiracEquation}
    (i\gamma^a e^{\mu}{}_a \overset{\rightarrow}{\nabla}_\mu -m)\psi=0\,,
\end{equation}
and its conjugate,
\begin{equation} \label{eq:CdiracEquation}
    \bar{\psi}(i\overset{\leftarrow}{\nabla}_\mu e^{\mu}{}_a \gamma^a+m)=0\,,
\end{equation}
where $e^{\mu}{}_a$ is the inverse tetrad. The differential equation of motion for $e^a$ is
\begin{equation}\label{eq:diraceomfore}
R^{\mu\nu}-\frac{1}{2}g^{\mu\nu}R=8\pi l_p^2\; T^{\mu\nu}\,,
\end{equation}
where
\begin{equation}
T^{\mu\nu} \equiv \frac{i}{2}\bar{\psi} \gamma^a{\overset{\leftrightarrow}{\nabla}}{}^{\nu} e^\mu{}_a \psi-\frac{1}{2}g^{\mu\nu}\bar{\psi} \Big(i\gamma^a\overset{\leftrightarrow}{\nabla}_\rho e^\rho{}_a-2m \Big)\psi\,,
\end{equation}
Here, we used the notation ${\overset{\leftrightarrow}{\nabla}}={\overset{\rightarrow}{\nabla}}-{\overset{\leftarrow}{\nabla}}$. As expected, the equations of motion (\ref{eq:DiracTorzijaJosJednom}), (\ref{eq:DiracEquation}), (\ref{eq:CdiracEquation}) and (\ref{eq:diraceomfore}) are precisely the equations of motion of the Einstein-Cartan-Dirac theory.

\subsection{\label{secIVsiv}Weyl and Majorana fields coupled to Einstein-Cartan gravity}

As is well known, the Dirac fermions are not an irreducible representation of the Lorentz group, and one can rewrite them as left-chiral and right-chiral irreducible Weyl fermion fields. Hence, it is useful to construct the $2$-crossed module and a constrained $3BF$ action for left and right Weyl spinors. For simplicity, we will discuss only the left-chiral spinor field (the right-chiral can be studied analogously). Additionally, we can also describe Majorana fermions using the same formalism, the only difference being the presence of an additional mass term in the Majorana action.

We soecify a $2$-crossed module $(L\stackrel{\delta}{\to} H \stackrel{\partial}{\to}G\,, \rhd\,, \{\_\,,\_\})$, in a way similar to the Dirac case, as follows. The groups are:
\begin{equation}
G=SO(3,1)\,, \quad \quad H=\mathbb{R}^4\,, \quad \quad L=\mathbb{R}^4(\grasmanovi)\,.
\end{equation}
The maps $\partial$,  $\delta$ and the Peiffer lifting are trivial. The action $\rhd$ of the group $G$ on $G$, $H$ and $L$ is given in the same way as for the Dirac case, whereas the spinorial representation reduces to
\begin{equation} \label{eq:actionOfGonLweyl}
M_{ab}\rhd P^{\alpha}=\frac{1}{2}(\sigma_{ab}){}^{\alpha}{}_{\beta} P^{\beta}\,, \qquad M_{ab} \rhd P_{\dot{\alpha}}=\frac{1}{2}(\bar{\sigma}{}_{ab}){}^{\dot{\beta}}{}_{\dot{\alpha}} P_{\dot{\beta}}\,,
\end{equation}
where $\sigma^{ab}=-\bar{\sigma}{}^{ab}=\frac{1}{4}(\sigma^a\bar{\sigma}^b-\sigma^b\bar{\sigma}^a)$, for $\sigma^a = (1, \vec{\sigma})$ and $\bar{\sigma}{}^a=(1,-\vec{\sigma})$, in which $\vec{\sigma}$ denotes the set of three Pauli matrices. The four generators of the group $L$ are denoted as $P^{\alpha}$ and $P_{\dot{\alpha}}$, where the Weyl indices $\alpha,\dot{\alpha}$ take values $1,2$.

The $3$-connection $(\alpha\,,\beta\,,\gamma)$ takes the form
\begin{equation}
\alpha = \omega^{ab}M_{ab}\,, \quad \quad \beta=\beta^a P_a\,, \quad \quad \gamma = \gamma_{\alpha} P^{\alpha}+ \bar{\gamma}{}^{\dot{\alpha}}P_{\dot{\alpha}}\,,
\end{equation}
while the $3$-curvature $(\cF\,, \cG\,, \cH)$ is
\begin{equation}
\begin{aligned}
     \cF = R^{ab} M_{ab}\,, \quad & \qquad \cG= \nabla \beta^a P_a\,,\\
     \quad \cH= \big(\D \gamma_{{\alpha}}+\frac{1}{2}\omega^{ab}(\sigma^{ab}){}^{{\beta}}{}_{\alpha}\gamma_{\beta}\big)P^{\alpha} & + \big( \D {\bar{\gamma}}{}^{\dot{\alpha}}+\frac{1}{2}\omega_{ab}({\bar{\sigma}}{}^{ab}){}^{\dot{\alpha}}{}_{\dot{\beta}}\bar{\gamma}^{\dot{\beta}}  \big) P{}_{\dot{\alpha}} \\
     \equiv (\overset{\rightarrow}{\nabla} \gamma){}_{\alpha} P^{\alpha} + ({\bar{\gamma}}\overset{\leftarrow}{\nabla}){}^{\dot{\alpha}}P{}_{\dot{\alpha}}\,. \hphantom{mm}&
\end{aligned}
\end{equation}
The Lagrange multiplier $D$ now contains as coefficients the spinor fields $\psi_{\alpha}$ and $\bar{\psi}^{\dot{\alpha}}$,
\begin{equation}
D = \psi_{\alpha} P^{\alpha} + \bar{\psi}^{\dot{\alpha}} P_{\dot{\alpha}}\,,
\end{equation}
and the bilinear form $\killing{\_}{\_}_{\mathfrak{l}}$ for the group $L$ is
\begin{equation}
  \begin{array}{c}
\killing{P^{\alpha}}{P^{\beta}}_{\mathfrak{l}} = \lc^{\alpha\beta}\,, \qquad
\killing{P_{\dot{\alpha}}}{P_{\dot{\beta}}}_{\mathfrak{l}} = \lc_{\dot{\alpha}\dot{\beta}}\,, \vphantom{\ds\int} \\
\killing{P^{\alpha}}{P_{\dot{\beta}}}_{\mathfrak{l}} = 0 \,, \qquad
\killing{P_{\dot{\alpha}}}{P^{\beta}}_{\mathfrak{l}} = 0 \,, \vphantom{\ds\int} \\
\end{array}
  \end{equation}
where $\lc^{\alpha\beta}$ and $\lc_{\dot{\alpha}\dot{\beta}}$ are the usual two-dimensional antisymmetric Levi-Civita symbols.

The pure $3BF$ action (\ref{eq:bfcgdh}) now becomes
\begin{equation}\label{eq:Weyltopoloski}
    S_{3BF}=\int_{\cM_4} B^{ab}\wedge R_{ab}+e_a\wedge\nabla\beta^a+ \psi{}^\alpha\wedge (\overset{\rightarrow}{\nabla} \gamma)_\alpha+ \bar{\psi}{}{}_{\dot{\alpha}}\wedge (\bar{\gamma} \overset{\leftarrow}{\nabla}){}^{\dot{\alpha}}\,.
\end{equation}
In order to obtain the suitable equations of motion for the Weyl spinors, we again introduce appropriate simplicity constraints, to obtain:
\begin{equation}\label{eq:Weyl}
\begin{aligned}
  S=&\int_{\cM_4} B^{ab}\wedge R_{ab}+e_a\wedge\nabla\beta^a+\psi{}^\alpha\wedge (\overset{\rightarrow}{\nabla} \gamma)_\alpha+ \bar{\psi}{}{}_{\dot{\alpha}}\wedge (\bar{\gamma} \overset{\leftarrow}{\nabla}){}^{\dot{\alpha}}\\
  &- \lambda_{ab} \wedge (B^{ab}-\frac{1}{16\pi l_p^2}\varepsilon^{abcd} e_c \wedge e_d)\vphantom{\ds\int}\\
  &-\lambda{}^\alpha\wedge ( \gamma{}_\alpha+\frac{i}{6} \varepsilon_{abcd}e^a\wedge e^b \wedge e^c\sigma^d{}_{\alpha\dot{\beta}}\bar{\psi}{}^{\dot{\beta}}) \vphantom{\ds\int} \\
  & -\bar{\lambda}{}_{\dot{\alpha}}\wedge(\bar{\gamma}{}^{\dot{\alpha}}+\frac{i}{6} \varepsilon_{abcd}e^a\wedge e^b \wedge e^c \bar{\sigma}{}^d{}^{\dot{\alpha}\beta}\psi_\beta)\vphantom{\ds\int}\\
  &-4\pi l_p^2 \varepsilon_{abcd} e^a\wedge e^b \wedge \beta^c (\bar{\psi}{}_{\dot{\alpha}}\bar{\sigma}{}^d{}^{\dot{\alpha}\beta}\psi_\beta)\,.\vphantom{\ds\int}
\end{aligned}
\end{equation}
The new simplicity constraints, in the third and fourth rows, feature the Lagrange multiplier $1$-forms $\lambda{}_\alpha$ and $\bar{\lambda}{}^{\dot{\alpha}}$. Also, in analogy to the coupling between the spin and the torsion in Einstein-Cartan-Dirac theory, the term in the fifth row is chosen to ensure that the coupling between the Weyl spin tensor
\begin{equation}
s_a \equiv i\varepsilon_{abcd} e^b \wedge e^c \;\psi^\alpha \sigma^d{}_{\alpha\dot{\beta}}\bar{\psi}{}^{\dot{\beta}}
\end{equation}
and torsion is given as:
\begin{equation} \label{WeylTorsionSpinCoupling}
T_a=4\pi l_p^2 s_a\,.
\end{equation}
The action for the Majorana field is precisely the same, but for an additional mass term in the action:
\begin{equation} \label{eq:MajoranaMassTerm}
-\frac{1}{12} m \varepsilon_{abcd}e^a\wedge e^b \wedge e^c  \wedge e^d ( \psi^\alpha  \psi_\alpha +\bar{\psi}_{\dot{\alpha}}\bar{\psi}^{\dot{\alpha}} ) \,.
\end{equation}

The variation of the action (\ref{eq:Weyl}) with respect to the variables $B_{ab}$, $\lambda^{ab}$, $\gamma{}_\alpha$, $\bar{\gamma}{}^{\dot{\alpha}}$, $\lambda{}_\alpha$, $\bar{\lambda}{}{}^{\dot{\alpha}}$, $\psi_\alpha$, $\bar{\psi}{}^{\dot{\alpha}}$, $e^a$, $\beta^a$ and $\omega^{ab}$ gives us the equations of motion, which can be algebraically solved for all variables as functions of $\psi_\alpha$, $\bar{\psi}{}^{\dot{\alpha}}$ and $e^a$:
\begin{equation} \label{eq:WeylLagMult}
\begin{aligned}
\beta^a{}_{\mu\nu} = 0\,, \quad \lambda^{ab}{}_{\mu\nu}=R^{ab}{}_{\mu\nu}\,,
&\quad
\lambda{}_\alpha{}_\mu=\nabla_\mu\psi_\alpha\,,
\quad 
\bar{\lambda}{}^{\dot{\alpha}}{}_\mu=\nabla_\mu \bar{\psi}{}^{\dot{\alpha}}\,,\vphantom{\ds\int}\\
B_{ab}{}_{\mu\nu}=\frac{1}{8\pi l_p^2}\varepsilon_{abcd} e^c{}_{\mu} e^d{}_{\nu}\,,
& \quad 
\omega_{ab\mu}=\triangle_{ab\mu}+K_{ab\mu}\,,\vphantom{\ds\int} \\
\gamma{}_\alpha{}_{\mu\nu\rho}=i\varepsilon_{abcd}e^a{}_\mu e^b{}_\nu e^c{}_\rho \sigma^d{}_{\alpha\dot{\beta}}\bar{\psi}{}^{\dot{\beta}}\,,
& \quad
\bar{\gamma}{}^{\dot{\alpha}}{}_{\mu\nu\rho}=i\varepsilon_{abcd}e^a{}_\mu e^b{}_\nu e^c{}_\rho \bar{\sigma}{}^d{}^{\dot{\alpha}\beta}\psi_\beta\,.
\end{aligned}
\end{equation}
In addition, one also obtains (\ref{WeylTorsionSpinCoupling}). Finally, the differential equations of motion for the spinor and tetrad fields are
\begin{equation}
\bar{\sigma}{}^{a\dot{\alpha}\beta} e^{\mu}{}_a \nabla_\mu \psi_\beta=0\,, \qquad
\sigma^a{}_{\alpha\dot{\beta}} e^{\mu}{}_a \nabla_\mu \bar{\psi}^{\dot{\beta}}=0\,,
\end{equation}
and
\begin{equation}
R^{\mu\nu}-\frac{1}{2}g^{\mu\nu}R=8\pi l_p^2 \; T^{\mu\nu}\,,
\end{equation}
where
\begin{equation}
\begin{array}{lcl}
  T^{\mu\nu} & \equiv & \ds \frac{i}{2}\bar{\psi}\bar{\sigma}{}^b e^{\nu}{}_b \nabla^\mu\psi+\frac{i}{2}\psi\sigma^b e^{\nu}{}_b \nabla^\mu\bar{\psi}  \vphantom{\ds\int} \\
 & & \ds -\frac{1}{2}g^{\mu\nu}\Big(i\bar{\psi}\bar{\sigma}^a e^{\lambda}{}_a \nabla_{\lambda} \psi+i\psi\sigma^a e^{\lambda}{}_a \nabla_{\lambda}\bar{\psi}\Big)\, . \vphantom{\ds\int} \\
\end{array}
\end{equation}
Here we have suppressed the spinor indices, for simplicity. In the case of the Majorana field, the equations of motion (\ref{eq:WeylLagMult}) remain the same. The equations of motion for $\psi_\alpha$ and $\bar{\psi}{}^{\dot{\alpha}}$ obtain the additional mass term,
\begin{equation}\label{eq:m12}
   i \sigma^a{}_{\alpha\dot{\beta}} e^{\mu}{}_a\nabla_\mu \bar{\psi}{}^{\dot{\beta}} -m\psi_\alpha =0\,, \qquad
  i \bar{\sigma}{}^{a\dot{\alpha}\beta} e^{\mu}{}_a \nabla_\mu \psi_\beta - m\bar{\psi}{}^{\dot{\alpha}} =0\,,
\end{equation}
while the stress-energy tensor becomes
\begin{equation}
\begin{array}{lcl}
T^{\mu\nu} & \equiv & \ds \frac{i}{2}\bar{\psi}\bar{\sigma}{}^b e^{\nu}{}_b \nabla^\mu\psi+\frac{i}{2}\psi\sigma^b e^{\nu}{}_b \nabla^\mu\bar{\psi} \vphantom{\ds\int} \\
 & & \ds - g^{\mu\nu}\frac{1}{2}\left[i\bar{\psi}\bar{\sigma}^a e^{\lambda}{}_a \nabla_{\lambda} \psi+i\psi\sigma^a e^{\lambda}{}_a \nabla_{\lambda}\bar{\psi}-\frac{1}{2}m\left(\psi\psi+\bar{\psi}\bar{\psi}\right)\right]\,. \\
\end{array}
\end{equation}

\section{\label{secV}Conclusions}

Let us summarize the results of the paper. In Section 2 we have introduced the $BF$ theory and discussed models based on constrained $BF$ action, in particular the Yang-Mills theory in Minkowski spacetime and the Plebanski formulation of general relativity. Section 3 was devoted to the first step in the categorical ladder and the $2BF$ theory. After introducing the notions of a $2$-group, a crossed module, and the corresponding $2BF$ theory, we have studied the $2BF$ formulation of general relativity and the Einstein-Yang-Mills theory. Then, in Section 4 we have performed one more step in the categorical ladder, and introduced the notions of a $3$-group, $2$-crossed module, and the $3BF$ theory. This structure was employed to construct the constrained $3BF$ actions for the cases of Klein-Gordon, Dirac, Weyl and Majorana fields, each coupled to the Einstein-Cartan gravity in the standard way. In those descriptions, it turned out that the scalar and fermion fields are associated to a {\em new gauge group}, similar to the gauge fields being associated to a gauge group in the Yang-Mills theory. This opens up a possibility of a classification of matter fields based on an algebraic structure of a $3$-group.

All the obtained results serve to complete the first step of the spinfoam quantization programme, as outlined in the Introduction. This paves the way to the study of steps 2 and 3 of the programme. Namely, the full action for gravity, gauge fields and matter is written completely in the langulage of differential forms, which can be easily adapted to a triangulated spacetime manifold, in the sense of Regge calculus. This can be seen in the following table:
{\footnotesize
\begin{center}
\setlength{\tabcolsep}{0pt}
    \label{tab:table4}
\begin{tabular}{|c|c|c|c|c|c|c|} \hline
\ \ $d$\ \ \  & \ \ triangulation$\vphantom{\ds\int}$\ \ \  &\ \  dual triangulation\ \ \  & \ \ form\ \ \   &\ \  fields\ \ \  &\ \ field strengths\ \ \  \\ \hline\hline
$0$ & vertex$\vphantom{\ds\int}$ & $4$-polytope & $0$-form & $\phi$, $\psi_{\tilde\alpha}$, $\bar{\psi}{}^{\tilde\alpha}$ & \\ \hline
$1$ & edge$\vphantom{\ds\int}$ & $3$-polyhedron & $1$-form & \ \  $\omega^{ab}$, $A^I$, $e^a$ \ \ \  &  \\ \hline
$2$ & triangle$\vphantom{\ds\int}$ & face & $2$-form & $\beta^a$, $B^{ab}$ & $R^{ab}$, $F^I$, $T^a$ \\ \hline
$3$ & tetrahedron$\vphantom{\ds\int}$ & edge & $3$-form & $\gamma$, $\gamma_{\tilde\alpha}$, $\bar\gamma{}^{\tilde\alpha}$  & $\cG^a$ \\ \hline
$4$ & $4$-simplex$\vphantom{\ds\int}$ & vertex & \ \  $4$-form\ \ \  & & $\cH$, $\cH_{\tilde\alpha}$, $\bar\cH{}^{\tilde\alpha}$ \\ \hline
\end{tabular}
\end{center}
}

This data can be utilized to construct a Regge-discretized topological $3BF$ action, and from that a state sum $Z$, giving rise to a rigorous definition of the path integral
\begin{equation} \label{StatementOfIntentWithMatter}
Z = \int \cD g \int \cD \phi \; e^{iS[g,\phi]}\,,
\end{equation}
which is a generalization of (\ref{StatementOfIntent}) in the sense that it adds matter fields (including the gauge boson sector) to gravity at the quantum level. Being a topological theory, and given the underlying structure of the $3$-group, a pure $3BF$ action ought to ensure the topological invariance of the state sum $Z$, i.e., $Z$ should be triangulation independent. This step, however, requires the generalizations of the Peter-Weyl and Plancharel theorems to $2$-groups and $3$-groups, which are unfortunately still missing (though there are some attempts to circumvent them at least in the $2$-group case~\cite{BaratinFreidel2015,DittrichEtAl2019}). Namely, the purpose of the Peter-Weyl and Plancharel theorems is to provide a decomposition of a function on a group into a sum over the corresponding irreducible representations, which then specifies the spectrum of labels for the simplices in the triangulation, and fixes the domain of values for the fields living on those simplices. In the absence of the two theorems, one can still try to {\em guess} the irreducible representations of the $2$- and $3$-groups, as was done for example in the {\em spincube model} of quantum gravity \cite{MikovicVojinovic2012}, or to try to construct the state sum using other techniques, as was done in~\cite{BaratinFreidel2015,DittrichEtAl2019}).

Of course, when building a realistic theory, we are not interested in a topological theory, but instead in one which contains local propagating degrees of freedom. Thus the state sum $Z$ need not be a topological invariant. This is obtained via the step 3 of the spinfoam quantization programme, by imposing the simplicity constraints on $Z$. The classical actions discussed in this paper manifestly distinguish the topological sector from the simplicity constraints, which have been explicitly determined. Imposing them should thus be a straightforward procedure for a given $Z$. Completing this programme would ultimately lead us to a tentative state sum describing both gravity and matter at a quantum level, which is a topic for future research.

In addition to the construction of a full quantum theory of gravity, there are also many additional possible studies of the classical constrained $3BF$ action. For example, a Hamiltonian analysis of the theory could be interesting for the canonical quantization programme, and some work has begun in this area \cite{RadenkovicVojinovicSymmetry2020}. Also, it is worth looking into the idea of imposing the simplicity constraints using a spontaneous symmetry breaking mechanism. Finally, one can also study in more depth the mathematical structure and properties of the simplicity constraints. The list is not conclusive, and there may be many other interesting topics to study.

\newpage

\ \ %

\end{document}